\documentclass[12pt]{article}

\usepackage{amssymb,amsmath,bm}
\usepackage{comment}
\usepackage{dsfont}

\usepackage{subcaption}
\usepackage{epsfig, epsf, graphicx}
\usepackage{pstricks, pst-node, psfrag}
\usepackage{color}
\usepackage{verbatim,enumerate}
\usepackage{rotating, lscape,natbib}
\usepackage{setspace}
\usepackage{tikz}
\usetikzlibrary{arrows,calc,tikzmark}
\usepackage{float}
\usepackage{caption}
\usepackage{natbib}

\usepackage{booktabs}
\usepackage{algorithm}
\usepackage{algpseudocode}

\newtheorem{prop}{Proposition}

\newtheorem{assumption}{Assumption}
\newtheorem{rem}{Remark}

\begin{document}
\thispagestyle{empty} \baselineskip=28pt \vskip 5mm
\begin{center} {\Huge{\bf Nonparametric Trend Estimation in Functional Time Series with Application to Annual Mortality Rates }}
	
\end{center}

\baselineskip=12pt \vskip 10mm

\begin{center}\large
%\if1\blind
%{
Israel Mart\'inez-Hern\'andez$^{*}$\footnote[1]{\label{note1} 
\baselineskip=10pt Statistics Program,
King Abdullah University of Science and Technology,
Thuwal 23955-6900, Saudi Arabia.\\
E-mail: israel.martinezhernandez@kaust.edu.sa, marc.genton@kaust.edu.sa\\
This research was supported by the
King Abdullah University of Science and Technology (KAUST).
 } and Marc G.~Genton$^{\ref{note1}}$
% } \fi
\end{center}

\baselineskip=17pt \vskip 10mm \centerline{\today} \vskip 15mm

\begin{center}
{\large{\bf Summary}}
\end{center}
Here, we address the problem of trend estimation for functional time series. Existing contributions either deal with detecting a functional trend or assuming a simple model. They consider neither the estimation of a general functional trend nor the analysis of functional time series with a functional trend component. Similarly to univariate time series, we propose an alternative methodology to analyze functional time series, taking into account  a functional trend component. We propose to estimate the functional trend by using a tensor product surface that is easy to implement,  to interpret, and allows to control the smoothness properties of the estimator. Through a Monte Carlo study, we simulate different scenarios of functional processes to show that our estimator accurately identifies the functional trend component. We also show that the dependency structure of the estimated stationary time series component is not significantly affected by the error approximation of the functional trend component. We apply our methodology to annual mortality rates in France. 

\baselineskip=14pt

\par\vfill\noindent
\par\medskip\noindent
%{\bf Short title}: Trend in Functional Time Series

\clearpage\pagebreak\newpage \pagenumbering{arabic}
\baselineskip=26pt

%%%%%%%%%%%
\section{Introduction}\label{sec:intro}

In many phenomena, data are collected on a large scale, resulting in high-dimensional and high-frequency data. This is why there has been an increasing amount of interest in functional data analysis (FDA). FDA deals with data, called functional data, that are defined on an intrinsically infinite-dimensional space. When the functional data are time-dependent, they are called functional time series. Some examples of data that can be considered as functional time series are the annual mortality rates and the annual temperature data. In practice, functional time series often tend to be nonstationary. This nonstationarity may be  caused by structural breaks, functional random walk components or deterministic trend components. Deterministic trends, or functional trends, can be observed in different phenomena where functional data approaches have been used, e.g., growth curves \citep{Ramsay-Silverman2005}, annual mortality rates \citep{Hyndmanetal2007}, gene networks \citep{telesca2009}, climate change \citep[][]{FraimanEtal2014}, electricity power systems \citep[][]{Horvathetal2015}, and EEG data \citep{hasenstab2017a}. The detection and estimation of the functional trend are crucial in data analysis, modeling and forecasting. 

The common method used to analyze functional time series involves projecting each curve on a finite dimensional space, for example, on the space generated by $r$ eigenfunctions, and then modeling the projected values by using multivariate time series techniques \citep[][]{Hyndmanetal2007, AueEtAl2015}. When the functional time series has a functional trend component, one can  still transform the curves into a vector and then model the trend component as in multivariate time series. However, using principal component  analysis to reduce dimensionality may not be appropriate,  since the estimation of the covariance operator is not consistent in this case. An alternative approach, similar to the univariate time series, is to estimate the functional trend directly from the functional data, then remove it, and  analyze the remaining functional time series. In this paper, we adopt the latter approach.

Functional trends are challenging because of the complexity of the space where functional data are defined. In multivariate time series, trends have only one component, i.e., they have the form $h(t)$, where $t$ represents time, and $h$ is a continuous function defined over time \citep[see for example][]{Wu&Zhao2007,Chen&Wu2018}. Unlike in multivariate time series, functional trends have an additional component: the continuous parameter of each functional data. That is, functional trends can be written as a function with two variables $T(s,t)$, where $s$ is the continuous parameter of each curve,  and $t$ represents time. 

\begin{figure}[!b]
\begin{center}
\includegraphics[scale=.58]{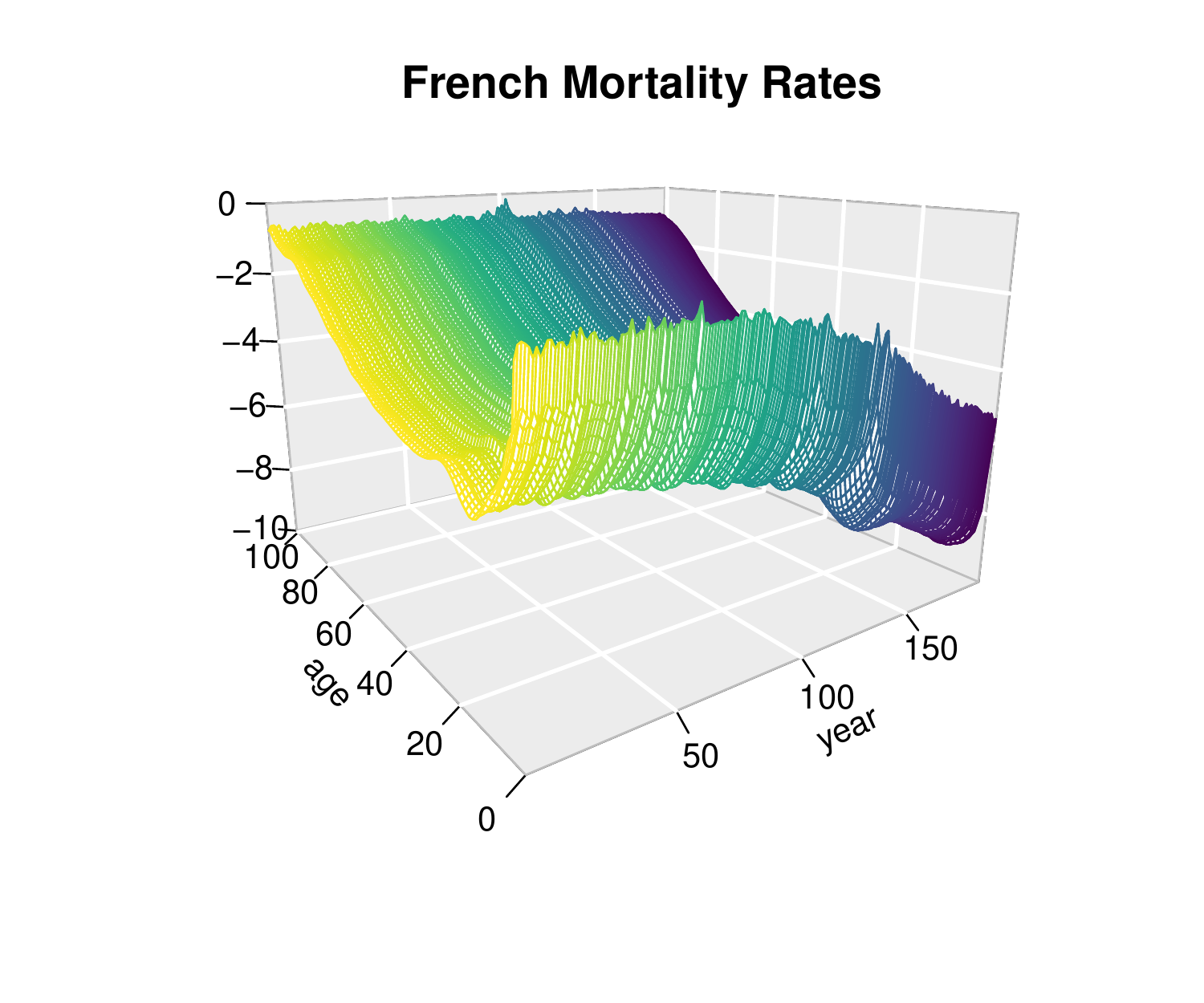} 
\includegraphics[scale=.58]{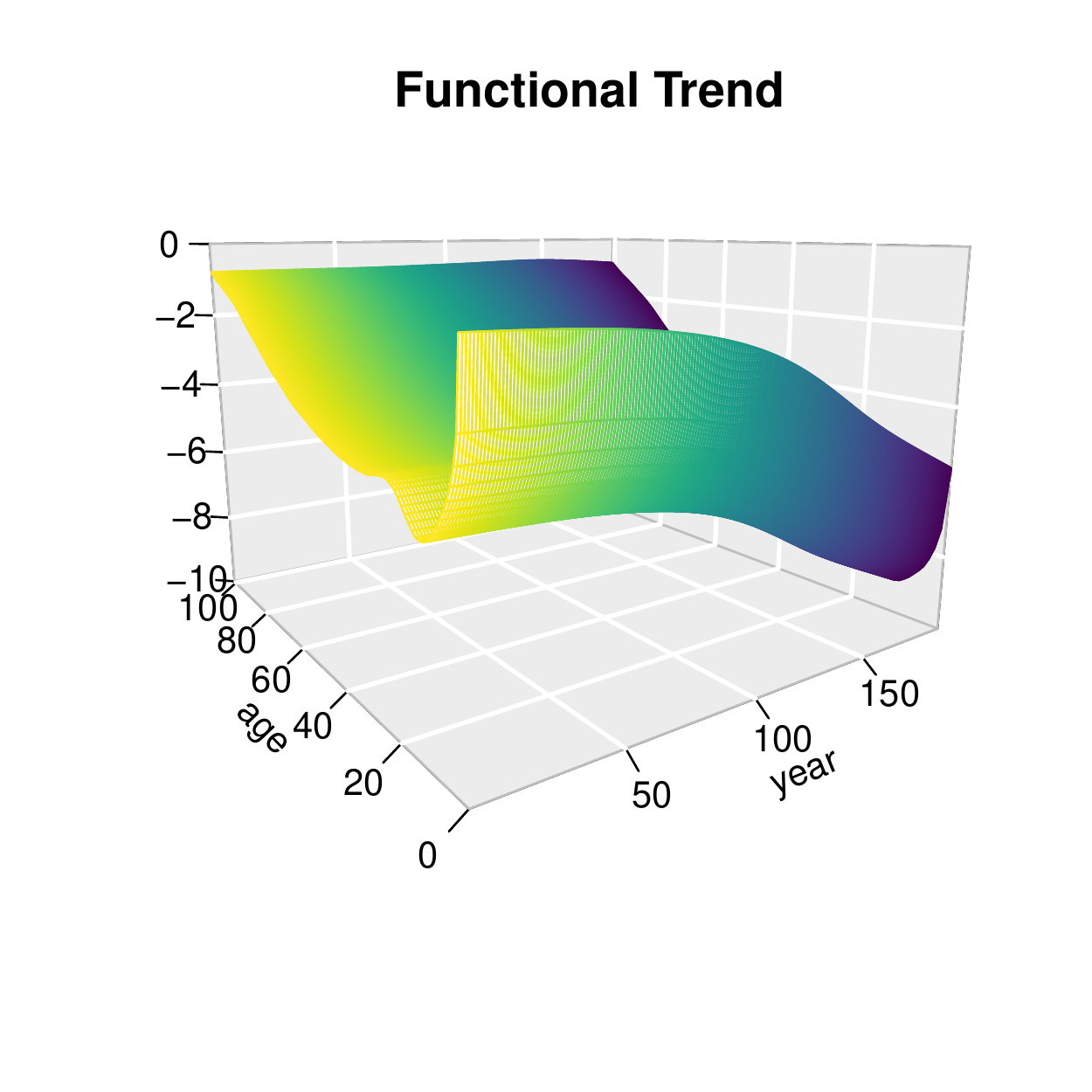} 
\caption{Functional time series of log mortality rates in France from 1816 to 2006, for zero to 100 years of age (left), and the corresponding  estimated functional trend (right). The estimated functional trend describes the smooth changes over time of the functional data.}
\label{Example1}
\end{center}
\end{figure}

A few attempts can be found in the literature on the study of functional trends.  In \cite{FraimanEtal2014} a functional trend is defined by using the concept of records, where a record means the occurrence of new extreme observations, but nothing is mentioned about the estimation. 
In \cite{KokoszkaYoung2017}, a hypothesis test of trend stationarity of functional time series was proposed. In that paper, the functional trend is assumed to be separable and linear in time, $T(s,t)= f(s)t$, and a least squares estimator is used to estimate $f(s)$. Although this may cover a large number of cases, which depend linearly on time, it is still a very specific model. Functional trend can take very complex shapes, e.g., 
 Figure \ref{Example1} shows  
log annual mortality rates in France from $1816$ to $2006$, where each point of $Y_{n}(s)$ represents the total mortality rate, in year $n$, at age $s\in [0,100]$. Across the years $n$, the log mortality rate has been decreasing for almost all ages $s$. For ages between $0$ and $60$, it seems that the decrease behaves  like a quadratic function, whereas for ages between $60$ and $100$, the values  behave like a  linear function. On the other hand, the $s$ coordinate (age) is dominated by a U-shaped curve for each $n$.  The right panel shows the resulting functional trend estimated by applying our proposed methodology. Here we analyze these data as a functional time series considering the functional trend $T(s,t)$ (Section \ref{DataAnalyses}).  Due to the complexity of functional trends, we propose describing $T(s,t)$ using a nonparametric  approach. 

The functional time series approach has several advantages over the multivariate time series methods. Multivariate methods ignore information about the underlying continuity behavior of the data. For example, the bivariate time series of the annual mortality rates at ages $s=40$ and $s=41$, $\{ Y_{n} (40), Y_{n}(41) \}^{\top}$, is permutable in the multivariate setting. This leads to a rough surface for a functional trend estimator. In contrast, smoothness is an important property of functional data. Thus, FDA extracts additional information contained in a continuous function or in its derivative \cite[][]{Kokoszka2012,UllahEtAl2013}. 

There is still a gap in knowledge on functional trends in functional time series. To the best of our knowledge, previous research either involved detecting functional trends or assuming a simple model, but none involved estimating a general functional trend nor  the analysis of functional time series with a functional trend component. Here, we describe a methodology to estimate the functional trend, and we show the analysis of the functional time series when the trend is taken into account.  We  propose  estimating a functional trend that is easy to implement and to interpret, and allows to control the smoothness properties of the estimator, which is useful in practice.

For instance, assume that $t$ is fixed in $T(s,t)$; thus $T(\cdot,t)$ can be interpreted as the ``common'' curve that  persists in different ways over time, weighted with the $t$ component. For example, if the  weight function is additive, i.e., $T(s,t)= f(s)+ g(t)$, then $f(s)$ can be considered as the mean curve and consequently the functional trend is simply $g(t)$. Now, if we fix $s\in D$, where $D$ represents the domain of the functional data, $T(s, \cdot)$ is the trend over time, and it can take different forms for each $s\in D$. Therefore, for each coordinate, $T(s,t)$ can take different shapes, and a nonparametric estimation for each coordinate seems reasonable. We propose using a  B-spline to describe the different forms for each coordinate. When the sample size tends to infinity, $T$ can be assumed to be continuous in $s$ and $t$, and resulting in a tensor product surface. To obtain the smoothness property of the tensor product B-spline, similar ideas from the univariate case \citep[][]{Eilers&Marx96} can be applied. One can opt to use one penalty parameter for both directions, or one for each direction, or a combination of both \citep[see][]{ Wood2003,XiaoEtal2013}. Here, we consider marginal penalizations as described in \cite{Wood2006}. This allows us to study  the trend over time and a possible trend within the domain $D$ separately. Also, this way of penalizing is easy to interpret and to control for each smoothness parameter.

The remainder of our paper is organized as follows. In Section \ref{sec:Trend}, we introduce the model that is assumed in this paper, and we develop the proposed estimator for the functional trend. In Section \ref{TP}, we study the theoretical properties of the proposed estimator, as well as the selection of the smoothing parameters. In Section \ref{sec:MC}, we conduct a simulation study to evaluate the performance of the proposed estimator under different simulation settings. In Section \ref{DataAnalyses}, we analyze a dataset of annual mortality rates assuming a functional trend component. Section \ref{D} presents some discussion. Proofs and additional material are provided in the Web Appendix.

\section{Trend in Functional Time Series}\label{sec:Trend}
\subsection{Preliminaries}

Assume that we observe a functional time series with sample size $N$, $\{Y_{1}, \ldots, Y_{N}\}$, taking values on a separable Hilbert space $\mathcal{H}$ that will be defined in Section \ref{MR}, i.e., $Y_{n}(s):D\to \mathbb{R}$ is a continuous function  for $n=1, \ldots, N$. Now, assume that $\{Y_{n}\}$ follows  the model 
\begin{equation} \label{ModelT}
Y_{n}(s) = T(s, n/N) + X_{n} (s), 
\end{equation}
where $T(s,t):D\times [0,1]\to \mathbb{R}$ is a deterministic function, and  $\{X_{n}\}$ is a stationary functional time series with $\mathbb{E}(X_{n})=0$. Thus, $\mathbb{E}(Y_{n})= T(s,n/N)$ and $\{Y_{n}\}$ is not weakly stationary. The function $T(s,t)$ is the trend component.

A technique that is widely used in time series to obtain the stationarity property is considering the first difference of $\{Y_{n}, \, n \geq 1\}$, i.e., $\Delta Y_{n}:= Y_{n} - Y_{n-1}$. If the functional time series has a random walk component or if it is a $I(1)$ functional process, $\{\Delta Y_{n}\}$ is stationary \citep[][]{BeareEtal2017}. However, if the nonstationary component is a deterministic function, as in model \eqref{ModelT}, the  transformation $\{\Delta Y_{n}\}$ does not guarantee to remove the trend component $T(s,t)$. Moreover $\Delta X_{n}$ might be nonstationary even though $\{X_{n}\}$ is stationary, and as a consequence  $\{\Delta Y_{n}\}$ might be nonstationary.  To clarify the above idea, assume for instance that $T(s,t)= \sin (2\pi t + s)$  in model \eqref{ModelT}. Thus, $T(s, \frac{n}{N}) - T(s, \frac{n-1}{N})$ depends on $n$, and then $ \Delta Y_{n}$ depends on $n$ as well. Therefore the estimation of  the functional trend $T(s,t)$ is necessary.

\subsection{Nonparametric functional trend estimator}\label{NFTE}
We observe that, for $n_{0}$ fixed in model \eqref{ModelT}, $Y_{n_{0}}(\cdot)= T(\cdot, n_{0}/N) + X_{n_{0}}(\cdot)$. Thus $T(\cdot, n_{0}/N)$ represents the mean curve of the functional data $Y_{n_{0}}$ at time $n_{0}$. If $s_{0}\in D$ is fixed, then $\{Y_{n}(s_{0}),\,  n=1, \ldots, N \}$ is a  univariate time series and $T(s_{0}, \cdot)$ represents the deterministic trend at $s_{0}$. In the latter case, $T(s_{0}, \cdot)$ can be obtained via nonparametric estimation, such as Nadaraya-Watson, local polynomial, wavelet, or spline methods.  Here we use the spline method, i.e., we assume that  $T(s_{0}, \cdot)=\sum_{i=1}^{k_{2}}b_{i}\eta_{i}(\cdot)=  \mathbf{b}^{\top} \boldsymbol{\eta}(\cdot)$, where $ \boldsymbol{\eta}^{\top}=( \eta_{1}, \ldots, \eta_{k_{2}} ) $  is a B-spline basis function defined on $[0,1]$. 

Similarly, one could repeat this procedure for a finite set of $s$ values and apply a multivariate time series technique. However, since $Y_{n}$ is assumed to be a continuous function in $s$, multivariate methods cannot be extended to functional data.  Multivariate methods ignore the continuity (smoothness) property of $Y_{n}$, that is,  $Y_{n}(s_{0})$ and $Y_{n}(s_{0}+\epsilon)$ are considered permutable for any $\epsilon >0$. In addition, these would involve estimating infinite parametric or nonparametric tendencies. Instead, we allow each coefficient $b_{i}$ to be a smooth continuous function of $s$, i.e., $T(s, \cdot)=\mathbf{b}^{\top}(s) \boldsymbol{\eta}(\cdot)$, and $b_{i}(s)$ can be modeled nonparametrically as well. Let $ \boldsymbol{\nu}^{\top}=( \nu_{1}, \ldots, \nu_{k_{1}} ) $ be another B-spline basis function defined on $D$, such that $b_{i}(s) = \sum_{j=1}^{k_{1}} \theta_{ji}\nu_{j}(s)$ for $i=1, \ldots, k_{2}$. Then, $T(s,t)$ can be written as 
\begin{equation}\label{Eq:Trend1}
T(s,t)= \sum_{j=1}^{k_{1}}  \sum_{i=1}^{k_{2}} \theta_{ji}\nu_{j}(s)\eta_{i}(t)=  \boldsymbol{\nu}^{\top}(s) \boldsymbol{\Theta} \boldsymbol{\eta}(t).
\end{equation}
We propose estimating the functional trend by using a tensor product of the two spaces $\mathrm{span}\{\nu_{1}, \ldots, \nu_{k_{1}}\}$ and $\mathrm{span}\{\eta_{1}, \ldots, \eta_{k_{2}}\}$.
To obtain smoothness properties of $T(s,t)$, we consider penalty terms associated with each coordinate \citep[][]{Wood2006}. That is, 
\begin{equation}\label{Penalty}
P(T) = \lambda_{1}\int_{[0,1]} (P_{1}T )(t) \mathrm{d}t  +  \lambda_{2}\int_{D} (P_{2}T )(s) \mathrm{d}s
\end{equation}
where $P_{1}T = \int \{ \frac{\partial^{2}}{\partial s^{2}} T(s,t) \}^{2} \mathrm{d}s$ and  $P_{2}T= \int \{ \frac{\partial^{2}}{\partial t^{2}} T(s,t) \}^{2} \mathrm{d}t$. Other quadratic penalties can be considered, such as $\int\int \{(LT)(t,s) \}^{2} \mathrm{d}s \mathrm{d}t$, with $L$ a linear operator (e.g., the Laplacian). Here, we adopt the marginal penalty \eqref{Penalty}, where $\lambda_{1}$ and $\lambda_{2}$ control    the smoothness of $T(s,t)$ in the first component and the second component, respectively. This penalty is invariant to a linear rescaling of the functional data, which is useful since, in practice the domain $D$ of the functions is rescaled to the interval $[0,1]$. Also, $P(T)$ is easily interpretable and allows us to control the smoothness in the direction of the  domain $D$ and in the direction of the time domain, separately, which is desirable for the estimation of the functional trend.

We observe that if $\lambda_{1}\gg 0 $, then $T(\cdot, n/N)$ is a linear function on $D$ for each $n=1,\ldots, N$, and if $\lambda_{1}=0$,  
then $T(\cdot, n/N)$ is close to the shape of the functional data $Y_{n}$, i.e., $T(\cdot, n/N)\approx Y_{n}$. Thus, to only capture the trend over time and without removing the inherent shape of the functional data, a $\lambda_{1}$ different from zero should be considered. Similarly, if  $\lambda_{2}\gg 0$, then $T(s, \cdot)$ represents a linear trend for each $s$, whereas when $\lambda_{2}=0$, then $T(s, \cdot)$ represents interpolation of $Y_{1}(s), \ldots, Y_{n}(s)$ for each $s$, and so $T(s,t)$ results in a rough surface. In Section \ref{SmoothSelection}, we describe how to select these parameters taking into account the dependency structure  of $\{X_{n}\}$. In practice, users are free to choose the values of $\lambda_{1}$ and $\lambda_{2}$, as well as the number of basis functions in each coordinate, $k_{1}$ and $k_{2}$. 

Given $P(T)$ we obtain the estimator of $T(s,t)$ by using a penalized least square estimator, that is, we obtain $\boldsymbol{\hat{\Theta}}$ minimizing the mean integrated squared error
\begin{equation}\label{Opt1}
\boldsymbol{\hat{\Theta}}=\arg \min_{\boldsymbol{\Theta}} \left[ \sum_{n=1}^{N} \int_{D} \{ Y_{n}(s) -  \boldsymbol{\nu}^{\top} (s) \boldsymbol{\Theta} \boldsymbol{\eta}(n/N) \}^{2}\mathrm{d}s + P(T) \right].
\end{equation}
Consequently, we define $\hat{T}(s,t)= \boldsymbol{\nu}^{\top} (s) \boldsymbol{\hat{\Theta}} \boldsymbol{\eta}(t)$.

In summary, we propose describing the deterministic trend in functional time series by using a smooth tensor product surface. A tensor product surface is very flexible in the sense that it can represent complex structures in functional data. Because of the penalization term, a few numbers of basis functions (or knots) are required, and it is computationally feasible. In Section~\ref{sec:MC}, we show the performance of our proposed estimator under different scenarios.

\subsection{Modeling with estimated functional trend}
Once the functional trend has been estimated, we make an $h$-step ahead forecast for the functional time series $\{Y_{n}\}$ by forecasting each component of the model \eqref{ModelT}, that is, $\hat{Y}_{N+h} = \hat{T}_{N+h} + \hat{X}_{N+h}.$ 
The $h$-step ahead forecast for each component is computed as follows: For the stationary functional time series component, we obtain $\hat{X}_{N+h}$ by modeling the functional time series $\{\tilde{X}_{n}:= Y_{n}(s)- \hat{T}(s, n/N) \}_{n=1}^{N}$. For example, one can use the methodology described in \cite{AueEtAl2015} (See Section \ref{Sec:DA}). To obtain the  $h$-step ahead forecast for the functional trend component, we use a Taylor expansion in the time direction. Specifically, we define the $1$-step ahead forecast as
\begin{equation}\label{Tforecast}
\hat{T}_{N+1}(s):= \hat{T}(s, 1) + \frac{1}{N+1}  \frac{\partial}{\partial t} T(s,t)\big|_{t=1},
\end{equation} 
where $ \hat{T}(s, 1)$ corresponds to the trend estimated at time $N$. This $1$-step ahead forecast is iterated $h$ times, with $\hat{T}(s, 1)$ being  the last trend observed or forecasted in each iteration. After the iterations, we obtain the $h$-step ahead forecast $\hat{T}_{N+h}$. In general,  $T(s,t)$ can be assumed to be a function with slow variation over time,  as evidenced in  Figure \ref{Example1}. Thus, in this paper we use the linear approximation \eqref{Tforecast}.

\section{Theoretical Properties}\label{TP}
The theoretical properties of penalized splines have been studied when errors are uncorrelated. For the one-dimensional setting, see, for example, \cite{Li&Rupperd2008}, and \cite{ClaeskensEtal2009}. Some papers that have studied the two-dimensional setting are \cite{Lai&Li2013} and \cite{Xiao2019}. \cite{Xiao2019} studied the asymptotic behavior of bivariate penalized tensor-product splines, extending the idea from the one-dimensional setting. Here, we adopt the same approach as in  \cite{Xiao2019} to study the consistency of the functional trend estimator $\hat{T}(s,t)$. 

Let $ \boldsymbol{ \mathcal{P}}_{1}$ and  $ \boldsymbol{ \mathcal{P}}_{2}$ be the fixed marginal penalty matrices, for the first component and the second component of $T(s,t)$, respectively. Thus, the first component of the penalty term in \eqref{Penalty} can be written as $\int (P_{1}T)(t)\mathrm{d}t= \int  \{ \boldsymbol{\Theta}\boldsymbol{\eta}(t) \}^{\top}   \boldsymbol{\mathcal{P}}_{1} \boldsymbol{\Theta}\boldsymbol{\eta}(t)\mathrm{d}t=\{\mathrm{vec}(  \boldsymbol{\Theta}) \}^{\top}  \boldsymbol{J_{\eta}}\otimes  \boldsymbol{ \mathcal{P}}_{1} \mathrm{vec}(  \boldsymbol{\Theta}),$ and the second component as $\int (P_{2}T)(s)\mathrm{d}s= \int \boldsymbol{\nu}^{\top}(s) \boldsymbol{\Theta} \boldsymbol{\mathcal{P}}_{2} \boldsymbol{\Theta}^{\top}\boldsymbol{\nu}(s)\mathrm{d}s=\{\mathrm{vec}(  \boldsymbol{\Theta}) \}^{\top} \boldsymbol{ \mathcal{P}}_{2}  \otimes \boldsymbol{J_{\nu}}  \mathrm{vec}(  \boldsymbol{\Theta}),$ where $\boldsymbol{J_{\nu}} = \int \boldsymbol{\nu}(s) \boldsymbol{\nu}^{\top}(s)\mathrm{d}s $, and $\boldsymbol{J_{\eta}} = \int \boldsymbol{\eta}(t) \boldsymbol{\eta}^{\top}(t)\mathrm{d}t $. Therefore 
\begin{equation}\label{PM}
P(T) = \{\mathrm{vec}(  \boldsymbol{\Theta}) \}^{\top}  \{  \lambda_{1} \boldsymbol{J_{\eta}}\otimes  \boldsymbol{ \mathcal{P}}_{1}  + \lambda_{2}   \boldsymbol{ \mathcal{P}}_{2}  \otimes \boldsymbol{J_{\nu}} \} \mathrm{vec}(  \boldsymbol{\Theta}),
\end{equation}
where  $\lambda_{1}$ and $\lambda_{2}$ are the smoothing parameters,
\begin{comment}
for the first component and the second component of $T(s,t)$, respectively
\end{comment}
and they need to be estimated.

\subsection{Functional representation}
We assume that the functional time series $Y_{n}(s)$ are given in the functional form. To establish the consistency of the functional trend estimator $\hat{T}(s,t)$, we introduce some concepts for functional time series. Let $\mathcal{H}$ be a Hilbert space of square integrable functions defined on a compact interval $D$, with inner product $\langle f,g \rangle = \int_{D} \! f(s)g(s) \mathrm{d}s$. Let $\{X_{n}(s), s\in D \}$ be a sequence of random variables in $\mathcal{H}$ with finite moments of order $2$, that is, for each $n$, $\mathbb{E}(\|X_{n} \|_{\mathcal{H}}^{2} ) < \infty$, where $\|\cdot \|_{\mathcal{H}}$ is the norm induced by the inner product in $\mathcal{H}$. Similarly to the univariate case, where the $\alpha$-mixing concept is required in the smoothing spline models with correlated random errors \citep[][]{Wang98}, one can assume short-range dependency in the functional time series $\{X_{n}\}$. We use the $L^{p}-m$-approximable concept (see supporting information for more details of this concept). 
	
Also, we will use the following assumptions that are concerned with the number of basis functions and the smoothing parameters.
\begin{assumption}\label{A1} $k_{1}k_{2}=o(N^{r})$ for some $r\in (0,1)$ and $\lim_{N \to \infty} k_{1}/ k_{2}= k_{0}$, for some constant $k_{0}$.
\end{assumption}
\begin{assumption}\label{A2} $\lambda_{1}k_{1}^{4 }=	O(N)$ and $\lambda_{2}k_{2}^{4 }=	O(N)$.
\end{assumption}	
\begin{assumption}\label{A3} The knots for the spline bases $\boldsymbol{\eta}$ and $\boldsymbol{\nu}$ are equidistantly distributed.
\end{assumption}
\begin{prop} \label{propFV}
Let $\{Y_{n}(s), s\in D\}$, for $n=1,\ldots, N$, be the functional time series observed, and following model \eqref{ModelT}. Suppose that $\{X_{n}\}$ is an $L^{4}-m$-approximable sequence, and that the functional trend has a tensor product representation $T(s,t)= \boldsymbol{\nu}^{\top} (s) \boldsymbol{ \Theta} \boldsymbol{\eta}(t) $ with $4$th-order derivatives. Then, under Assumptions \ref{A1}, \ref{A2}, and \ref{A3},
$$\mathbb{E}\left\{ \|\hat{T}(s,t)-  T(s,t) \|^{2}_{L_{2}}\right\} = o(1).$$
\end{prop}

In a nonparametric regression estimation, the long-run covariance of the time series plays an important role when errors are correlated. The assumption of $\{X_{n}\}$ being an $L^{4}-m$-approximable sequence implies that the corresponding long-run covariance operator is convergent. 

\begin{rem}  The weak dependence condition on $\{X_{n}\}$ is across time, $n$. Thus, for each $n$, $\{X(s):=X_{n}(s), s\in D\}$ can be a nonstationary process. This is another advantage of FDA over the multivariate methods. 
\end{rem}

\subsection{Matrix representation}\label{MR}
In practice, we do not observe continuous curves. Instead, each functional data $Y_{n}(s)$ is observed on a grid of points $\mathbf{s}_{n}=\{s_{n1}, \ldots, s_{nm}\}$. Without loss of generality, let us assume identical grids $\mathbf{s}_{n}\equiv \mathbf{s}= \{s_{1}, \ldots, s_{m}\}$ for $n=1, \ldots, N$. 
 Let $\mathbf{V}= \{ \boldsymbol{ \nu}(\mathbf{s}) \}^{\top}$ be the $m\times k_{1}$-matrix of the evaluation of $k_{1}$ basis functions on $m$ locations $\mathbf{s}$, let $\mathbf{Z}=\{ \boldsymbol{ \eta}(\mathbf{t}) \}^{\top}$ be the $N\times k_{2}$-matrix of the evaluation of $k_{2}$ basis functions on $N$ times $\mathbf{t}= \{1/N, 2/N, \ldots, 1\}$, and let $\mathbf{Y}= \{ Y_{1}(\mathbf{s}), Y_{1}(\mathbf{s}), \ldots, Y_{N}(\mathbf{s}) \}^{\top}$ be the $m\times N$-matrix of the observed functional time series, where each column represents observations of each continuous curve. Then, considering \eqref{Eq:Trend1},  model \eqref{ModelT} can be written as 
\begin{equation}\label{ModelTM}
\mathbf{Y}= \mathbf{V} \boldsymbol{\Theta} \mathbf{Z}^{\top} + \mathbf{X},
\end{equation}
 where  $\mathbf{X}$ denotes  the $m\times N$-matrix representing the evaluation of the functional time series $X_{n}(s)$ at $\mathbf{s}$, for $n=1, \ldots, N$.

 Thus, by using \eqref{PM}, the optimization problem \eqref{Opt1} is equivalent to 
 $
  \arg \min_{\boldsymbol{\Theta}} \| \mathbf{Y} -\mathbf{V} \boldsymbol{\Theta} \mathbf{Z}^{\top} \|^{2} +  \{\mathrm{vec}(  \boldsymbol{\Theta}) \}^{\top}  \{  \lambda_{1} \boldsymbol{J_{\eta}}\otimes  \boldsymbol{ \mathcal{P}}_{1}  + \lambda_{2}   \boldsymbol{ \mathcal{P}}_{2}  \otimes \boldsymbol{J_{\nu}} \} \mathrm{vec}(  \boldsymbol{\Theta}),
 $
 where $\| \cdot \|$ is the Frobenius norm, i.e., $\| \mathbf{E} \|= (\sum \sum | e_{ij} |^{2} )$ if $\mathbf{E}=(e_{ij})$. Thus, the solution $\boldsymbol{\hat{\Theta}}$  for $\boldsymbol{\Theta} $ satisfies,
 \begin{equation}\label{Eq:SolM}
 \left[ (\mathbf{Z}\otimes \mathbf{V} )^{\top} (\mathbf{Z}\otimes \mathbf{V} ) +  \lambda_{1} \boldsymbol{J_{\eta}}\otimes  \boldsymbol{ \mathcal{P}}_{1}  + \lambda_{2}   \boldsymbol{ \mathcal{P}}_{2}  \otimes \boldsymbol{J_{\nu}}   \right] \mathrm{vec} (\boldsymbol{\hat{\Theta}})= (\mathbf{Z}\otimes \mathbf{V} )^{\top} \mathrm{vec} (\mathbf{Y}).
 \end{equation}
Then, given $\lambda_{1}$ and $\lambda_{2}$, equation \eqref{Eq:SolM} can be solved with the \texttt{smooth.bibasis} function in the \textit{fda} R  package.  

\begin{prop} \label{Prop1} Assume that the functional time series is observed in a matrix form $\{Y_{n}(s_{i})\}$, $n=1\ldots, N$, $i=1,\ldots,m$, on a regular grid $ \mathbf{s}= \{s_{1}, \ldots, s_{m}\}$, and follows model \eqref{ModelT}. Suppose that $\{X_{n}\}$ is an $L^{4}-m$-approximable sequence, and that the functional trend has a tensor product representation $T(s,t)= \boldsymbol{\nu}^{\top} (s) \boldsymbol{ \Theta} \boldsymbol{\eta}(t) $ with $4$th-order derivatives. Then, under Assumptions \ref{A1}, \ref{A2}, and \ref{A3}, with the sample size $Nm$, 
$$\mathbb{E} \left\{ \|\hat{T}(s,t)-  T(s,t) \|^{2}_{L_{2}} \right\}= o(1),$$
where $\hat{T}(s,t) =  \boldsymbol{\nu}^{\top} (s)\hat{\boldsymbol{ \Theta}} \boldsymbol{\eta}(t)$, and $\hat{\boldsymbol{ \Theta}}$  is the solution of equation \eqref{Eq:SolM}.
\end{prop}

\begin{rem}
If each curve of the functional time series is observed on an irregular or sparse grid, we can always write model \eqref{ModelT} in a matrix form as in \eqref{ModelTM}, with $\mathbf{V}$ and $\mathbf{Z}$ matrices evaluated on the corresponding grids. 
\end{rem}

\subsection{Smoothing parameters selection}\label{SmoothSelection}
When considering penalized regression splines, the number of basis functions $k_{1}$ and $k_{2}$ (or knots) do not have a significant influence on the resulting penalized fit \citep[][]{Ruppert2002}. Usually, the number of basis functions grows with the sample size, but at a slower rate. Thus, the selection of $\lambda_{1}$ and $\lambda_{2}$ is more crucial, since these parameters control the flexibility of the tensor product. One of the advantages of tensor product surfaces is that all methods for curves are generalized easily. In particular, the methods to estimate the smoothing parameter can be extended to surfaces, such as Cross-Validation (CV),  Generalized Cross-Validation (GCV) or Akaike information criterion (AIC). In \cite{Wood2006}, the GCV method is used to estimate the smoothing parameters $\lambda_{1}$ and $\lambda_{2}$. While these methods perform well for uncorrelated errors, they perform poorly with  correlated errors, tending to underestimate (or overestimate) the smoothing parameters. In general, nonparametric estimators are sensitive to the presence of correlation in the errors, and several methods have been proposed. In \cite{OpsomerEtal2001}, one can find a general review of the literature in kernel regression, smoothing splines, and wavelet regression under correlated errors.

One possible solution to the correlated error problem is using a linear mixed effect model to represent the spline model. For instance,  assume that the functional time series $\{Y_{n}\}$ follows a Gaussian process. Thus, $ \mathrm{vec} (\mathbf{Y}) $ is a vector with Gaussian distribution, and $\mathrm{vec} (\boldsymbol{\Theta})$ can be estimated from the penalized log-likelihood function. Let $\boldsymbol{\hat{\Theta}}_{\mathrm{ML}}$ be the estimator obtained from the penalized log-likelihood function. If the vector $\mathrm{vec} (\mathbf{X})$  in model \eqref{ModelTM} has each entry being an independent random variable, then $\boldsymbol{\hat{\Theta}}_{\mathrm{ML}}$ satisfies equation \eqref{Eq:SolM}. Since the penalized tensor product in \eqref{ModelTM} can be considered as a linear mixed effect model, the estimator  $\boldsymbol{\hat{\Theta}}_{\mathrm{ML}}$ results in the posterior Bayes estimate (or best linear unbiased predictor). The latter has the advantage that the smoothing parameters $\lambda_{1}$ and $\lambda_{2}$ can be selected by using restricted maximum likelihood (REML). Moreover, in \cite{KrivobokovaEtal2007} it is shown that the selection of the smoothing parameters based on REML is robust under correlation structures. Based on these observations, we propose using the REML to select $\lambda_{1}$ and $\lambda_{2}$ under the assumption of independent residuals and a Gaussian distribution, i.e., $ \mathrm{vec} (\mathbf{X})\sim N(\mathbf{0}, \sigma^{2}_{X} \mathbf{I}_{mN})$ with $\mathbf{I}_{mN}$ as the identity matrix. Although the  extension is straightforward for surfaces, it is computationally expensive. Since the penalty \eqref{Penalty} of $T(s,t)$ is for each coordinate separately, by taking into account the average on the other coordinate, we propose using the REML on  the marginal mean data instead of using the whole dataset. With our proposal, the computational time is drastically reduced without losing the accuracy of the estimator. 
 
Specifically, we estimate $\lambda_{1}$ by using REML with the empirical mean $\frac{1}{N}\sum_{n=1}^{N} Y_{n}$ of the observed functional time series at $\mathbf{s}$.  Similarly, we estimate $\lambda_{2}$ by using the univariate time series $\{ \int Y_{1}(s) \mathrm{d}s , \ldots, \int Y_{N}(s) \mathrm{d}s\}$. To gain an intuition about this estimation, see the supporting information.  Once we have estimated $\lambda_{1}$ and $\lambda_{2}$ we solve \eqref{Eq:SolM} to obtain $\hat{\boldsymbol{\Theta} }$.
\begin{rem}
The estimated smoothing parameter $\hat{\lambda}_{1}$ controls the shape of the mean curve of the functional time series $\{Y_{n}\}$. On the other hand, the mean curve represents the common shape of the functional data over time. Thus, $\hat{T}(s,t)$ is expected to represent the shape in the $s$ coordinate. The estimated smoothing parameter $\hat{\lambda}_{2}$ represents the shape of the trend of  the average data in each period $n=1, \dots,N$. That is, $\hat{\lambda}_{2}$ controls the common trend of the functional time series, and so, $\hat{T}(s,t)$ represents the shape of the functional trend. 
\end{rem}

The methodology proposed here to estimate the functional trend in functional time series is easy to implement and computationally efficient. To obtain the estimators $\hat{\lambda}_{1}$ and  $\hat{\lambda}_{2}$, we can use the \texttt{gam} function in  the \textit{mgcv} package. Given $\hat{\lambda}_{1}$ and  $\hat{\lambda}_{2}$, we can obtain $\hat{\boldsymbol{\Theta} }$ using the \texttt{smooth.bibasis} function in the \textit{fda} package. An R code example of this implementation is included in the supporting information.

\section{Numerical Properties} \label{sec:MC}
\subsection{Preliminaries}

We investigate the performance of our proposed method under different scenarios. We use the \texttt{gam} function combined with the \texttt{smooth.bibasis} function in the \textit{mgcv} and \textit{fda} packages \citep[][]{fda}, respectively. To the best of our knowledge, there is no paper addressing functional trend estimation when functional data are observed over time. 

In the literature of nonparametric models, we can find methods related to the estimation of $T(s,t)$. However, these methods assume that the residuals $\{ X_{n}(s_{i})\}$ are independent (or uncorrelated) for all $n=1,\ldots,N$ and $i=1, \ldots, m$. Notice that our method does not require independence or stationarity of $\{ X_{n}(s_{i})\}_{i=1}^{m}$.

We denote our method as $T_{\mathrm{TPS}}$, where TPS stand for  tensor product surface. We compare it with the following estimators:
\begin{enumerate}
\item Finite Element Method: $\hat{T}_{\mathrm{FEM}}(s,t) := \sum_{j=1}^{k} \hat{a}_{j} \psi_{j} (s,t)$, where $\psi_{j}$ is a quadratic basis function associated with each node defined at points where data are observed, i.e., $(s_{i}, t_{n})$. The coefficients $\hat{a}_{j}$ are obtained  using a penalized least square method. Details of this method can be found in \cite{AzzimonteEtAl2015}. For this method, we use the \textit{fdaPDE} package \citep[][]{fdaPDE} to obtain $\hat{T}_{\mathrm{FEM}}(s,t)$.    
\item Kernel method: $\hat{T}_{\mathrm{Ker}}(s,t) := \frac{ \sum_{n=1}^{N} Y_{n} (s)K \left\{ \frac{ (s,t)- (s,n/N) }{h} \right\}  }{  \sum_{n=1}^{N} K \left\{ \frac{(s,t)-(s,n/N)}{h} \right\}  } $, where $K ( s,t  )= \frac{1}{2\pi } \exp \{-\frac{ s^{2} + t^{2}}{2} \}$. The bandwidth $h$ is selected via cross validation.
\item Linear trend: $\hat{T}_{\mathrm{Lin}}(s,t):=  \hat{\mu}(s)+ t \hat{f}(s)$, where $\hat{\mu}(s)= \bar{Y}_{n}(s) - \hat{f}(s)\frac{N+1}{2} $, and  $
\hat{f}(s)= \frac{1}{s_{N}} \sum_{n=1}^{N} \left(n - \frac{N+1}{2} \right) Y_{n}(s)$  with  $s_{N}= \sum_{n=1}^{N} \left( n - \frac{N+1}{2} \right)^{2}.$
\item\label{Naive} Naive method: For each $s_{i}\in \{s_{1},\ldots, s_{m}\}$, $\hat{T}_{\mathrm{Naiv}}(s_{i}, t):= \sum_{j=1}^{k_{2}} \hat{\theta}_{j} (s_{i}) \eta_{j} (t),$ where $(\eta_{1}, \ldots, \eta_{k_{2}})$ is a B-spline basis function. The coefficients $ \hat{\theta}_{j}$ are obtained via a penalized  least squares method, and using the \textit{fda} package \citep[][]{fda}. The penalty term is selected via generalized cross validation.
\item\label{Sand}  Sandwich smoother:  $\hat{T}_{\mathrm{Sand}}(s,t) := \sum_{j=1}^{k_{1}} \sum_{i=1}^{k_{2}} \hat{a}_{j,i} \nu_{j}(s) \eta_{i}(t)$. This estimator has the same form as \eqref{Eq:Trend1}, but the smoothing method is different \citep[See][]{XiaoEtal2013}. The sandwich smoother is implemented in the \texttt{fbps} command in the \textit{refund} package \citep[][]{refund}, and the corresponding smoothing parameters are selected via generalized cross validation. 
\item\label{ThinP}  Thin plate regression splines: $\hat{T}_{\mathrm{ThinP}}(s,t):= \sum_{j=1}^{k_{1}} \sum_{i=1}^{k_{2}} \hat{b}_{j,i} \nu_{j}(s) \eta_{i}(t)$. This estimator is a  tensor product as well with a  thin plate energy penalty \citep[See][]{Wood2003}. This method is implemented in the \texttt{gam} command in the \textit{mgcv} package. The corresponding smoothing parameter is selected via generalized cross validation.
\end{enumerate}

The estimator $\hat{T}_{\mathrm{FEM}}(s,t)$ is commonly used in cases where the domain is complex. In our case, the domain is a simple rectangle.  $\hat{T}_{\mathrm{Lin}}(s,t)$ is a basic parametric linear trend model,  and we use it as a baseline to measure the accuracy of our method in the simplest case (linear). $\hat{T}_{\mathrm{Naiv}}(s,t)$ is the most commonly used estimator on fMRI data to detrend time series at each voxel separately, see for example \cite{TANABE2002902} and \cite{HernandoEtAl2017}. $\hat{T}_{\mathrm{Sand}}(s,t)$ and $\hat{T}_{\mathrm{ThinP}}(s,t)$ are smooth tensor product surfaces proposed in nonparametric regression.

\subsection{Simulation setting}
We simulate $\{Y_{n}(s); \, s\in [0,1],\, n=1, \ldots, N \}$ from model \eqref{ModelT} with six different functional trends, $T(s,t)$, defined as follows: (a) $T_{1}(s,t)= 2 s +30  t$, (b) $T_{2}(s,t)= 25 t \sin(2 \pi s) $, (c) $T_{3}(s,t)= 20 t^{2} - 5 t + 5$, (d) $T_{4}(s,t)= 2(0.5 s + 4 t)^{2}$,  (e) $T_{5}(s,t)= 28 \sin (2 \pi t + s)$, and (f) $T_{6}(s,t)= \frac{4.5}{\pi \sigma_{s} \sigma_{t}} \exp \left\{ \frac{-(s- 0.2)^{2}}{\sigma_{s}^{2}} - \frac{(t-0.3)^{2}}{\sigma_{t}^{2}} \right\} +  \frac{2.7}{\pi \sigma_{s} \sigma_{t}} \exp \left\{ \frac{-(s- 0.7)^{2}}{\sigma_{s}^{2}} - \frac{(t-0.8)^{2}}{\sigma_{t}^{2}} \right\},$
with $\sigma_{s}= 0.3$ and $\sigma_{t}=0.4$. The function $T_{6}(s,t)$ was used in \cite{Wood2003} and in \cite{XiaoEtal2013} to study the performance of $\hat{T}_{\mathrm{ThinP}}(s,t)$ and $\hat{T}_{\mathrm{Sand}}(s,t)$, respectively.  The resulting surfaces for each of these models can be visualized in the supporting information.  

The stationary functional time series component, $\{X_{n}\}$, is simulated from the functional autoregressive model of order one (FAR$(1)$), defined as $X_{n}(s)= C_{1}\int_{[0,1]} \! \beta(u,s) X_{n-1}(u) \mathrm{d}u + W_{n}(s)$, with kernel $\beta(u,v)= \exp\{- (u^{2} + v^{2})/2 \}$, and functional white noise $\{W_{n}\}$ as  independent Brownian motion defined in $[0,1]$, where the scalar $C_{1}$ is such that the norm of the corresponding coefficient operator is $0.7$, that is, $\{ \int_{[0,1]} \int_{[0,1]}\! \beta^{2}(u,v) \mathrm{d}u  \mathrm{d}v \}^{1/2}= 0.7$.  We consider different sample sizes $N=100,300$ and $500$. For each $n=1, \ldots, N$, we simulate $Y_{n}(s)$ on an equispaced $50$-point  grid on $[0,1]$. Each simulation set is replicated $1000$ times. 

For each simulation we compute the functional trend. For our method $T_{\mathrm{TPS}}$ and for methods \ref{Naive}, \ref{Sand}, and \ref{ThinP},  we fix $k_{1}=10$ and $k_{2}=15$ in all cases. To compare the performance of our estimator $\hat{T}_{\mathrm{TPS}}$ with the competitors, we consider two different criteria. 

First, we evaluate the accuracy of the estimation of the functional trend component, computing the corresponding Integrated Squared Error ($\mathrm{ISE}_{T}$) defined as
$\mathrm{ISE}_{T}^{2}= \int_{[0,1]}\int_{[0,1]} \! \{T(s,t)- \hat{T}(s,t) \}^{2} \mathrm{d}s\mathrm{d}t.$
Second, we evaluate the accuracy of the estimation of the kernel $\beta(u,v)$ after removing the estimated functional trend. To do this, we estimate the kernel $\beta$ from the residual functional time series $\{\tilde{X}_{n}(s) \}= \{Y_{n}(s)- \hat{T}(s,n/N)\}$. We denote this estimator by $\hat{\beta}_{Y}$. Since our goal is not to have the best estimator of the kernel $\beta$, we assume that $\hat{\beta}_{X}$ is the truth, where $\hat{\beta}_{X}$ is the estimator  obtained from the original simulated functional time series $\{X_{n}\}$. Thus, we compare the estimator $\hat{\beta}_{Y}$ with the estimator $\hat{\beta}_{X}$ by computing the corresponding Integrated Squared Error ($\mathrm{ISE}_{\beta}$) defined as
$\mathrm{ISE}_{\beta}^{2}= \int_{D}\int_{D} \! \{ \hat{\beta}_{X}(s,t)- \hat{\beta}_{Y}(s,t) \}^{2} \mathrm{d}s\mathrm{d}t .$
The kernel estimators $\hat{\beta}_{Y}$ and $\hat{\beta}_{X}$ are obtained by using the \texttt{linmod} function with $15$ B-spline basis functions for each coordinate $u$ and $v$. Other parameters required in the  \texttt{linmod} function are set to be equal in both cases, $\hat{\beta}_{Y}$ and $\hat{\beta}_{X}$, to make them comparable. 

The value $\mathrm{ISE}_{T}$ represents the error approximation of the functional trend, while $\mathrm{ISE}_{\beta}$ indicates the difference between $\{Y_{n}(s)- \hat{T}(s, n/N) \}$ and $\{X_{n}\}$ in terms of dependency structure over time. Thus, $\mathrm{ISE}_{\beta}$ can be interpreted as the error dependency structure between $\{Y_{n}(s)- \hat{T}(s, n/N) \}$ and $\{X_{n}\}$ that is caused by the error approximation $T(s,t)- \hat{T}(s,t)$ of the functional trend.

\subsection{Simulation results} \label{SimResult}

\begin{figure}[!b]
\begin{center}
\includegraphics[scale=.45]{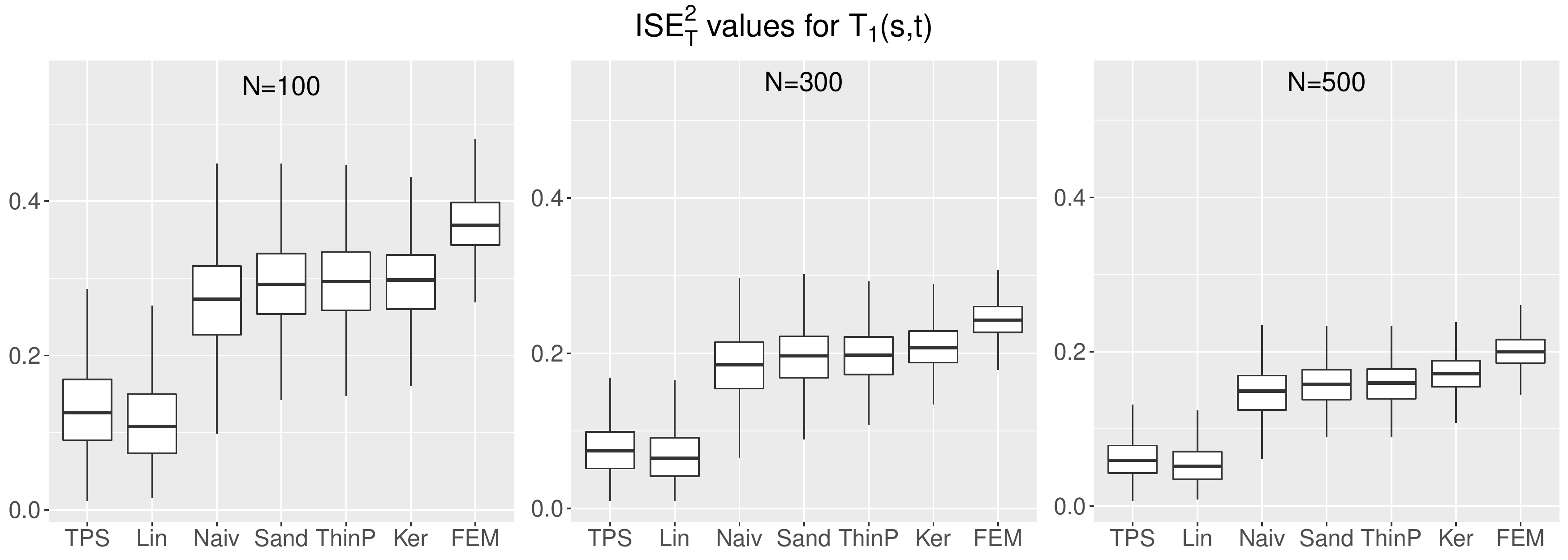} 
\vspace{0.2cm}

\includegraphics[scale=.45]{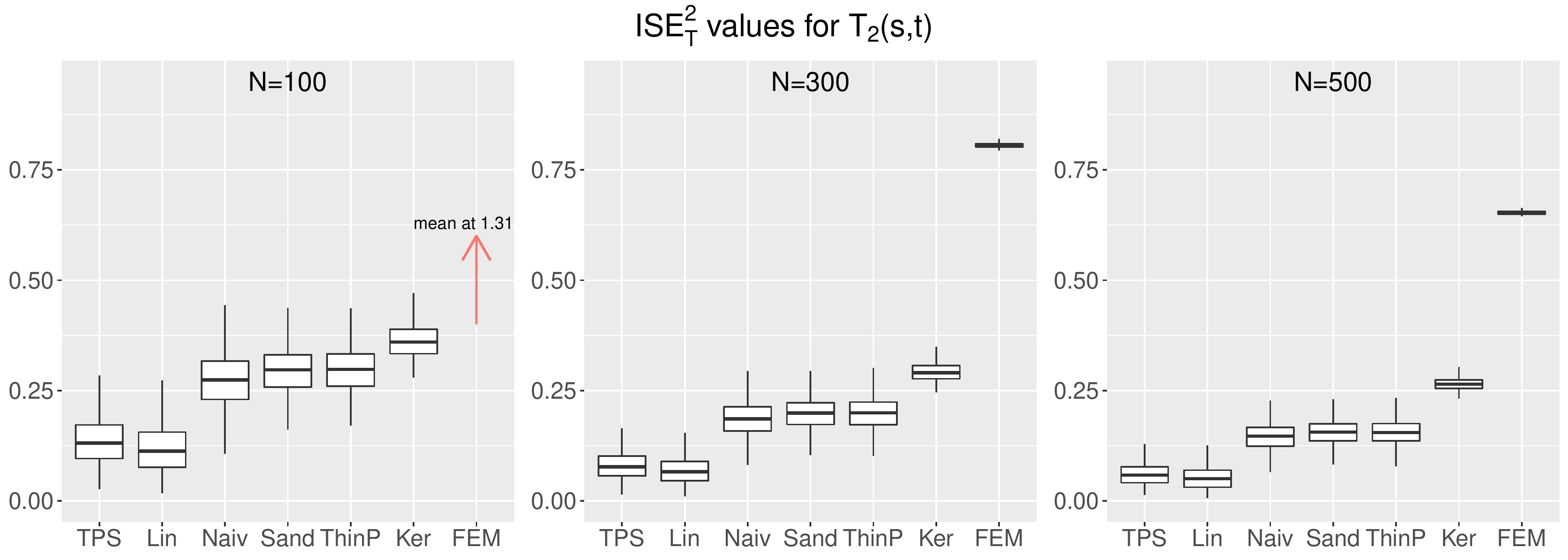} 
\caption{Boxplots of the $\mathrm{ISE}_{T}^{2}$ values for each simulation $\{Y_{n}, n=1,\ldots, N\}$ with functional trends $T_{1}$ and $T_{2}$, 
and different sample sizes  $N=100,300$ and $500$. A red arrow indicates that the $\mathrm{ISE}_{T}^{2}$ values are out of visual range and its mean is reported. Our proposed estimator $\hat{T}_{\mathrm{TPS}}$ and $\hat{T}_{\mathrm{Lin}}$ outperform the others.}\label{ISET12}
\end{center}
\end{figure}

\begin{figure}[!t]
\begin{center}
\includegraphics[scale=.45]{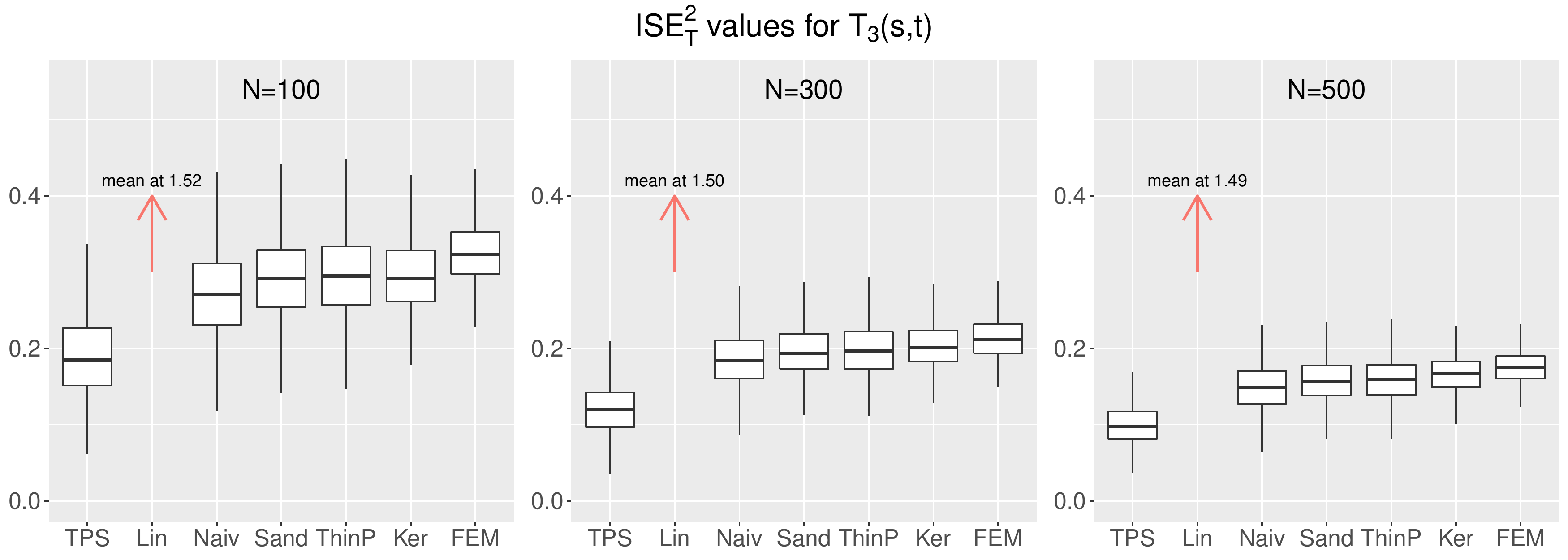} 
\vspace{0.2cm}

\includegraphics[scale=.45]{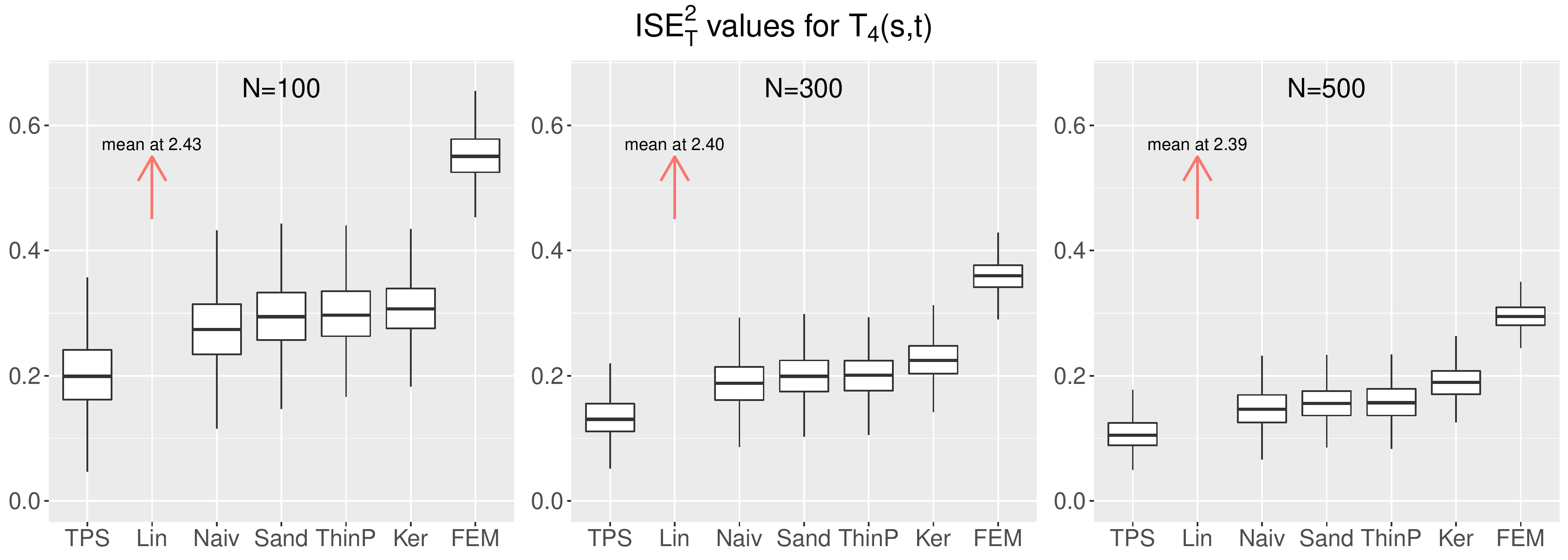} 
\caption{Boxplots of the $\mathrm{ISE}_{T}^{2}$ values for each simulation $\{Y_{n}, n=1,\ldots, N\}$ with functional trends $T_{3}$ and $T_{4}$ and different sample sizes $N=100,300$ and $500$. A red arrow indicates that the $\mathrm{ISE}_{T}^{2}$ values are out of visual range and its mean is reported. Our proposed estimator $\hat{T}_{\mathrm{TPS}}$ outperforms the others.}\label{ISET34}
\end{center}
\end{figure}

\begin{figure}[!t]
\begin{center}
\includegraphics[scale=.45]{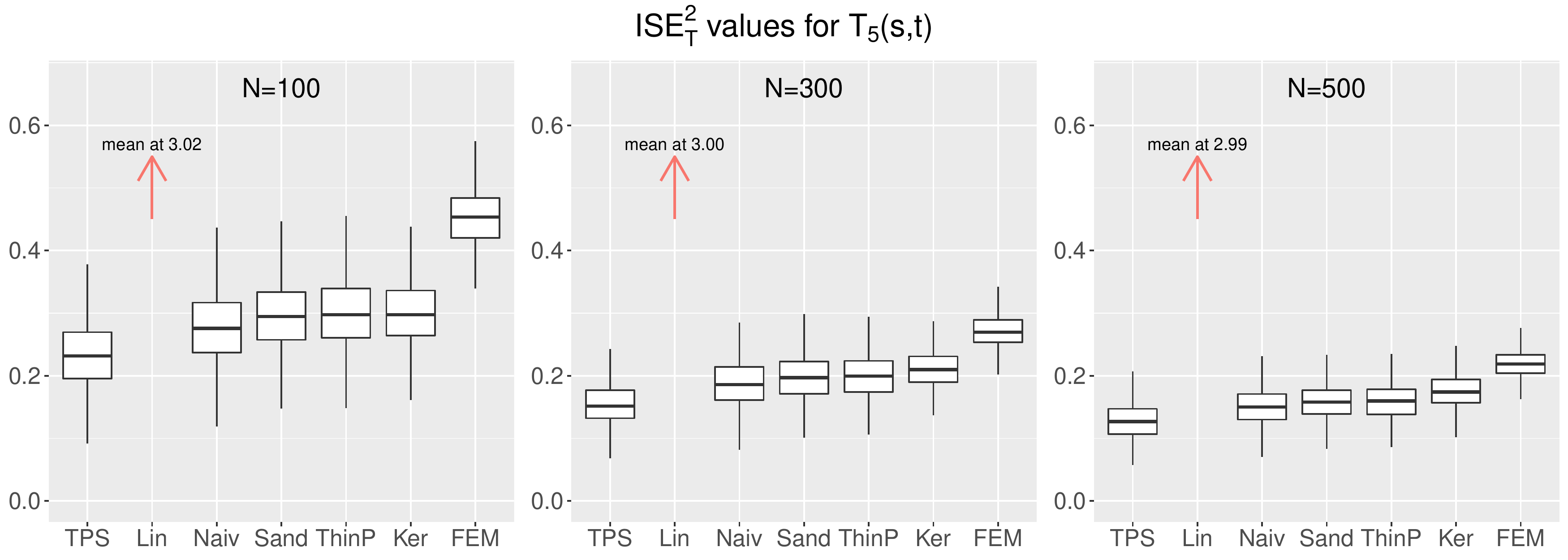} 
\vspace{0.2cm}

\includegraphics[scale=.45]{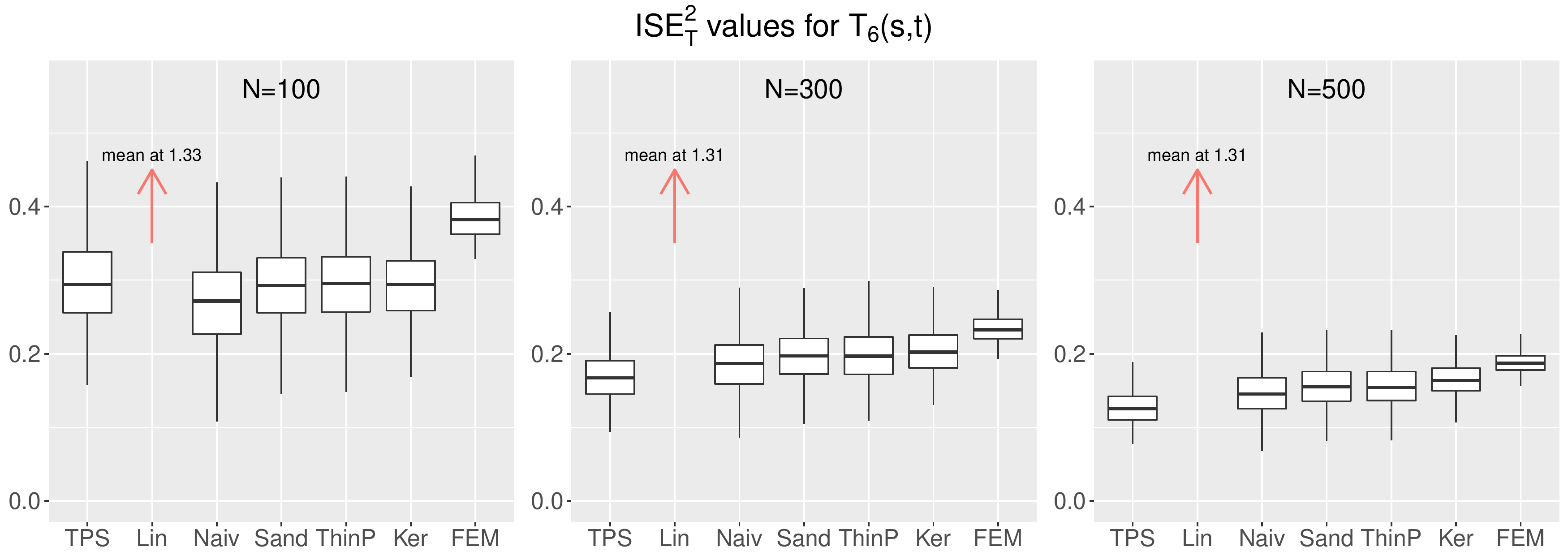} 
\caption{Boxplots of the $\mathrm{ISE}_{T}^{2}$ values for each simulation $\{Y_{n}, n=1,\ldots, N\}$ with functional trends $T_{5}$ and $T_{6}$, 
and different sample sizes $N=100,300$ and $500$. A red arrow indicates that the $\mathrm{ISE}_{T}^{2}$ values are out of visual range and its mean is reported. Our estimator $\hat{T}_{\mathrm{TPS}}$ has a good performance in all cases.}\label{ISET56}
\end{center}
\end{figure}

We present the results according to the shape of the functional trends over time: linear (Figure \ref{ISET12}), quadratic (Figure \ref{ISET34}), and  complex (Figure \ref{ISET56}). 

Figure \ref{ISET12} shows that our estimator $\hat{T}_{\mathrm{TPS}}$ and $\hat{T}_{\mathrm{Lin}}$ are highly accurate for the linear  functional trends $T_{1}$ and $T_{2}$. Both estimators have the lowest error values and they decrease when the sample size increases. Thus, in these cases, our proposed estimator performs as well as the parametric estimator $\hat{T}_{\mathrm{Lin}}$, with the advantage that our estimator does not require the specification of the functional trend shape. The results are similar for the functional trends $T_{3}$ and $T_{4}$ (Figure \ref{ISET34}). The $\mathrm{ISE}_{T}$ values for $\hat{T}_{\mathrm{TPS}}$ remain as accurate as in the linear trends, except that  the $\mathrm{ISE}_{T}$ values for $\hat{T}_{\mathrm{Lin}}$ become significantly larger,  which is expected since the functional trends are not linear anymore. Therefore, our proposed  estimator outperforms the rest of the estimators on the quadratic functional trend. The latter conclusion extends to the $T_{5}$ and $T_{6}$ functional trends. Also, we observe that, in the case of nonlinear trends, the $\hat{T}_{\mathrm{Naiv}}$ estimator is the second best estimator after our method.

Next, we analyze the $\mathrm{ISE}_{\beta}$ values that represent the errors of the dependency structure caused by the error approximation of the functional trend estimator. We only present results corresponding to sample size $N=300$. The results from $N=100$ and $N=500$ are similar. Boxplots of all cases can be found in the supporting information. Table \ref{tableBetaN300}  shows the corresponding mean values and the standard deviations in parenthesis.  We observe that the $\mathrm{ISE}_{\beta}$ values behave similarly to the $\mathrm{ISE}_{T}$ values in almost all cases of different functional trends, except for the trend $T_{6}$. The $\mathrm{ISE}_{\beta}$ values are similar for $\hat{T}_{\mathrm{TPS}}$ and $\hat{T}_{\mathrm{Lin}}$ when considering functional trends $T_{1}$ and $T_{2}$. For  $T_{3}$ and $T_{4}$, the $\mathrm{ISE}_{\beta}$ values are significantly larger with the competitor estimators, whereas, for the $\hat{T}_{\mathrm{TPS}}$ estimator, the $\mathrm{ISE}_{\beta}$ values remain small. The conclusion is the same for the functional trend $T_{5}$. For  $T_{6}$, surprisingly, the estimator $\hat{T}_{\mathrm{FEM}}$ presents the lowest mean value of $\mathrm{ISE}_{\beta}$. However, $\hat{T}_{\mathrm{FEM}}$ performs poorly in all cases  when approximating the functional trend, i.e., $\hat{T}_{\mathrm{FEM}}$ presents the largest  $\mathrm{ISE}_{T}$ values.

In general, we conclude that  our proposed estimator performs well in all cases, even with simple models such as models $T_{1}$ and $T_{2}$ of the functional trend. It has the advantage of being applicable to a general class of functional trends with complex structures, and accurately describes the functional trends. 

\begin{table} \centering
%\scriptsize
\caption{Mean of the $\mathrm{ISE}_{\beta}^{2}$ values for each simulation $\{Y_{n}, n=1,\ldots, N\}$ with different functional trends, $T_{i}(s,t)$, and sample size $N=300$. Bold font is used to highlight the best performance. The corresponding standard deviations are indicated in parenthesis.} 
\begin{tabular}{lcccccc}
\toprule \toprule
 & $T_{1}(s,t)$  & $T_{2}(s,t)$  & $T_{3}(s,t)$  &   $T_{4}(s,t)$  &   $T_{5}(s,t)$ &  $T_{6}(s,t)$  \\   \toprule
 %$N=300$              &   	            & 			   & 			 &  			&  		&     \\
 \mbox{TPS}  &0.028 (0.02) & 0.030 (0.02) & $\mathbf{0.057}$ (0.03)& $\mathbf{0.076}$ (0.04)& $\mathbf{0.114}$ (0.05)& 0.371  (0.09)\\
 \mbox{Lin}       & $\mathbf{0.014}$ (0.01)& $\mathbf{0.015}$ (0.02)& 0.452 (0.07) & 0.464  (0.07)& 1.071 (0.07)& 0.955 (0.05)\\
 \mbox{Naiv}    & 0.151 (0.06)& 0.154 (0.06)& 0.154 (0.06)& 0.158 (0.06)& 0.153 (0.06)& 0.153 (0.06)\\
 \mbox{Sand}   & 0.157 (0.06) & 0.159 (0.06) & 0.159 (0.06) & 0.162 (0.06) & 0.157 (0.06) & 0.158 (0.06) \\
 \mbox{ThinP}  & 0.165 (0.06) & 0.169 (0.06)& 0.168 (0.06)& 0.173 (0.06)& 0.166 (0.06)& 0.168 (0.06) \\
 \mbox{Ker}      & 0.263 (0.06)& 0.320 (0.07)& 0.276 (0.06)& 0.209 (0.06)& 0.251 (0.06)& 0.297 (0.07) \\
 \mbox{FEM}    & 0.080 (0.04)& 0.432 (0.04) & 0.068 (0.03)& 0.221 (0.06)& 0.160 (0.04)& $\mathbf{0.145}$ (0.03)\\
 \toprule
\end{tabular}
\label{tableBetaN300}
\end{table}  

\section{Data Analysis}\label{DataAnalyses}\label{Sec:DA}
\subsection{Objectives}
In this section, we apply our methodology on annual mortality rates in France. Our goal is to show that the consideration of a functional trend from a functional point of view improves data analysis, in particular data forecasting. We model the dataset considering the functional trend described in Section \ref{NFTE}.  Then, we compare the forecasted with the model without considering the functional trend.

To forecast functional time series, we adopt one of the most feasible and commonly used procedures. Let $\{Z_{n}(s), n=1, \ldots, N\}$ be a functional time series with sample size $N$. For each $n$, $Z_{n}$ is transformed into a vector time series of dimension $r$, $\mathbf{Z}_{n}= (z_{n,1}, \ldots, z_{n,r} )^{\top}$, by projecting $Z_{n}$ into $r$ functional principal components. Then, the multivariate time series $\{\mathbf{Z}_{n}, n=1, \ldots, N\}$ is modeled by using VAR$(p)$ or ARIMA models. Using the fitted time series model, and for $h$ fixed, we obtain the $h$-step ahead forecast $\hat{\mathbf{Z}}_{N+h}= (\hat{z}_{N+h,1}, \ldots, \hat{z}_{N+h,r} )^{\top} $. Finally, we multiply the predicted vector $\mathbf{\hat{Z}}_{N+h}$ by the $r$ estimated principal components to obtain the $h$-step ahead forecast of functional time series $\hat{Z}_{N+h}(s)$  \citep[see][ for more details]{Hyndmanetal2007, AueEtAl2015}. Here,  we model  each component of $\{\mathbf{Z}_{n}\}$ separately, similarly as in \cite{Hyndmanetal2007}. 

Thus, to see the differences between considering and not considering the functional trend $T(s,t)$, we apply the latter methodology 
described in the functional time series $\{Y_{n}, n=1, \ldots, N\}$, and in the functional time series $\{\tilde{X}_{n}, n=1, \ldots, N\}$, where $\tilde{X}_{n}(s):= Y_{n}(s)- \hat{T}(s,n/N) $ and $\hat{T}(s,n/N)$ is obtained as described in Section \ref{sec:Trend}. The corresponding models for the univariate time series are selected with  Akaike information criterion (AIC). 

\subsection{Mortality rates in France}

\begin{figure}

\begin{subfigure}[t]{1\textwidth}
\begin{center}
    \includegraphics[scale=.42]{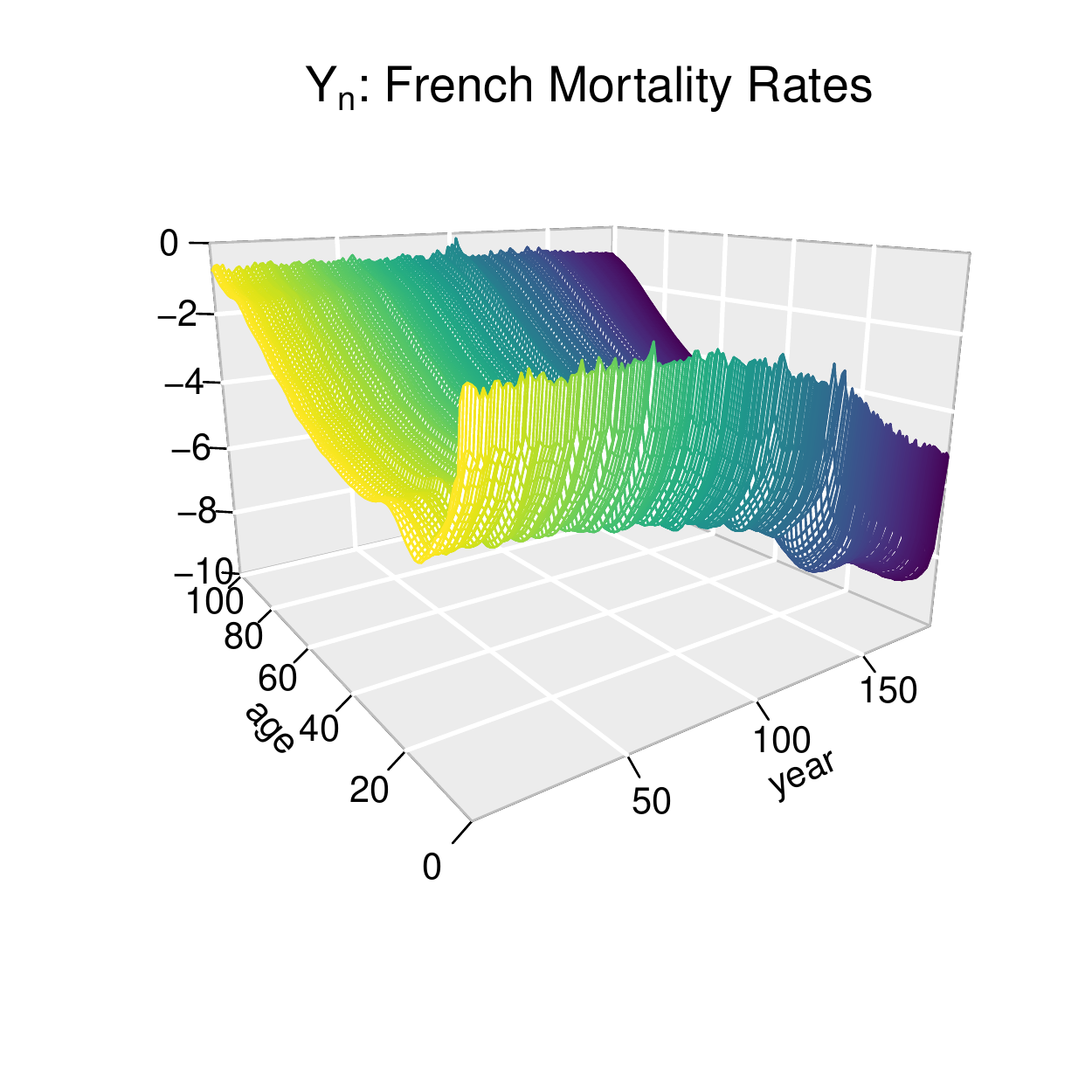} 
    \includegraphics[scale=.42]{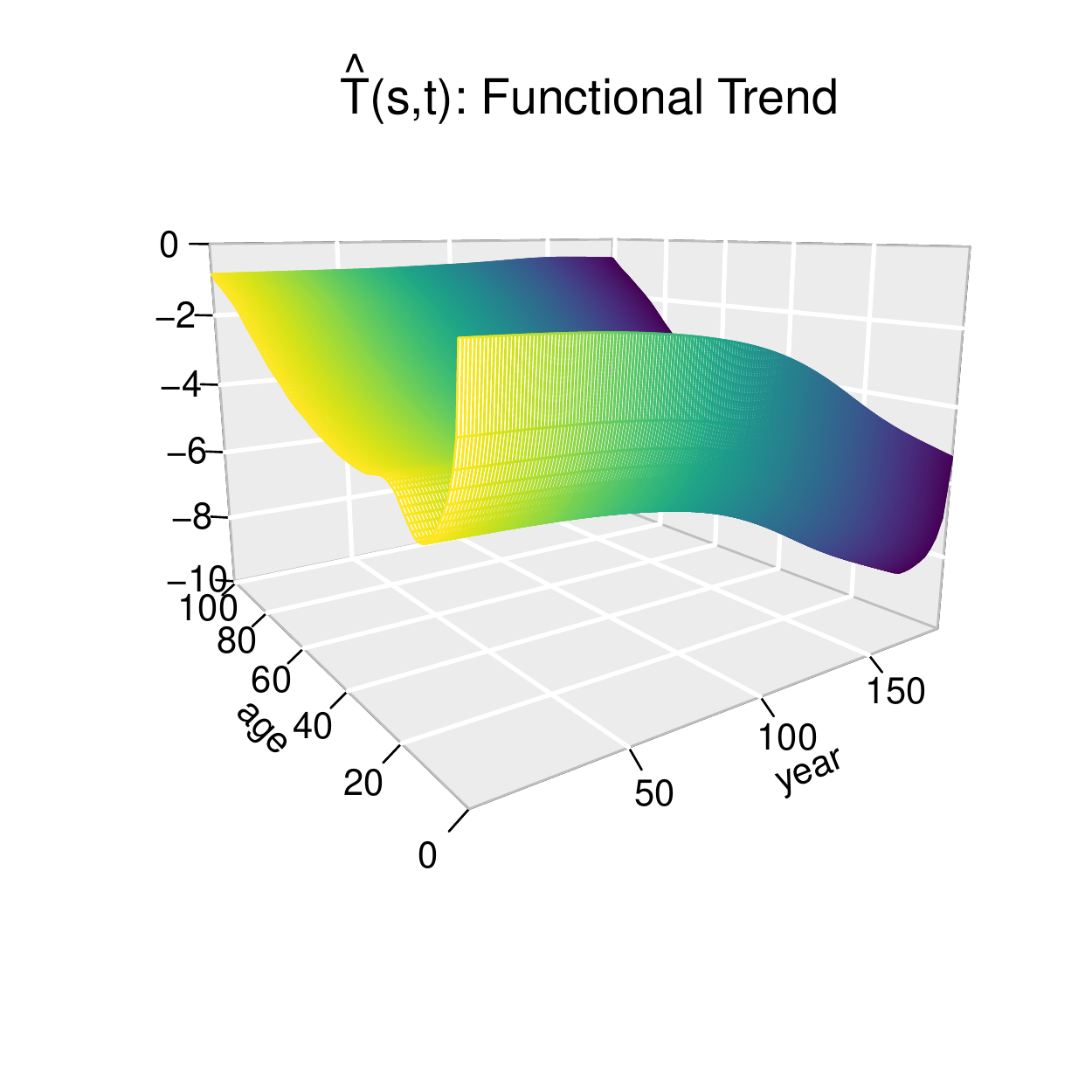} 
    \includegraphics[scale=.42]{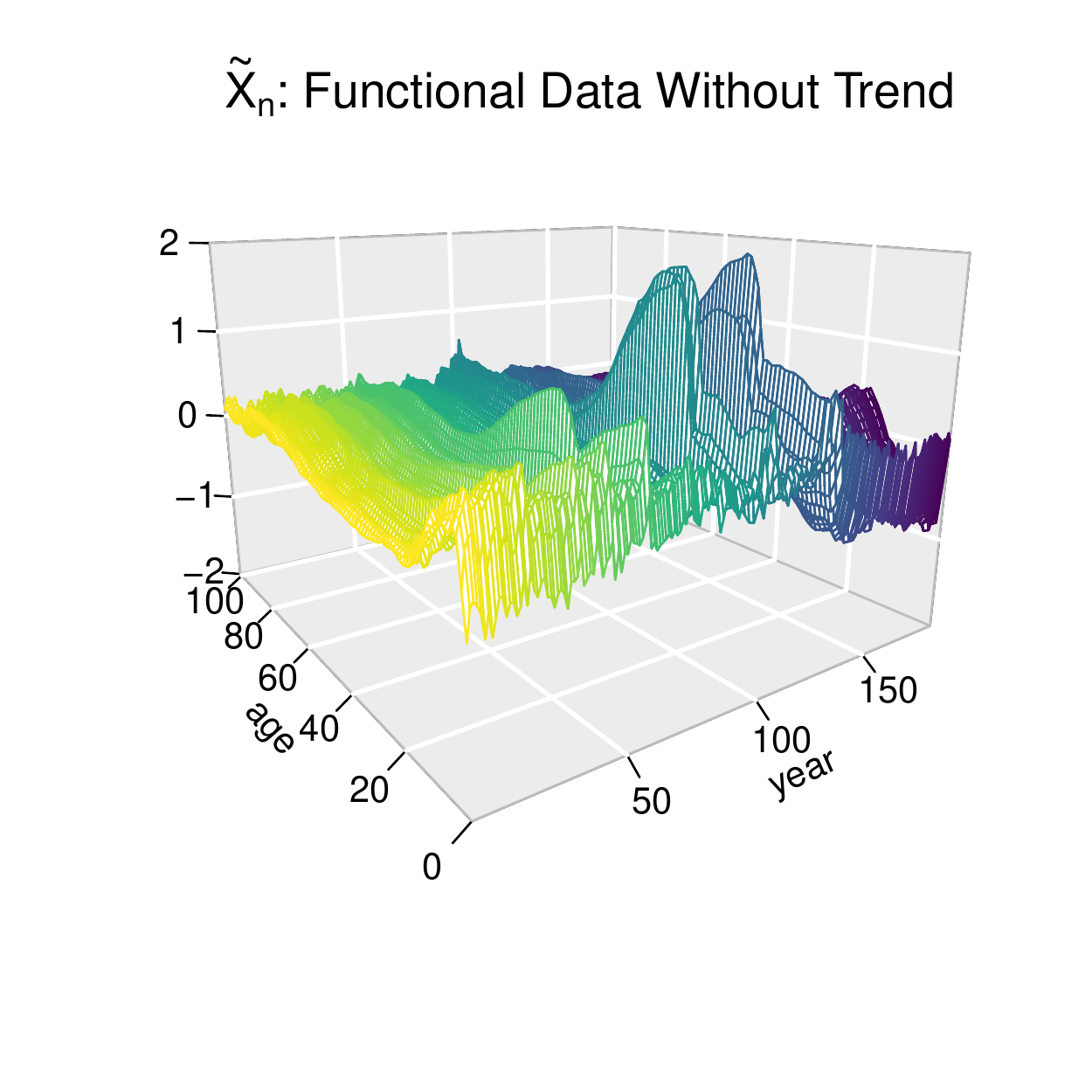} 
    \caption{Functional data $\{Y_{n}\}$ observed (left).  Estimated functional trend (center), and functional data after removing the estimated functional trend (right).}\label{Data1}
      \end{center}
  \end{subfigure}  \\
  
  \bigskip
  
  \begin{subfigure}{1\textwidth}
  \begin{center}
  \includegraphics[scale=.6]{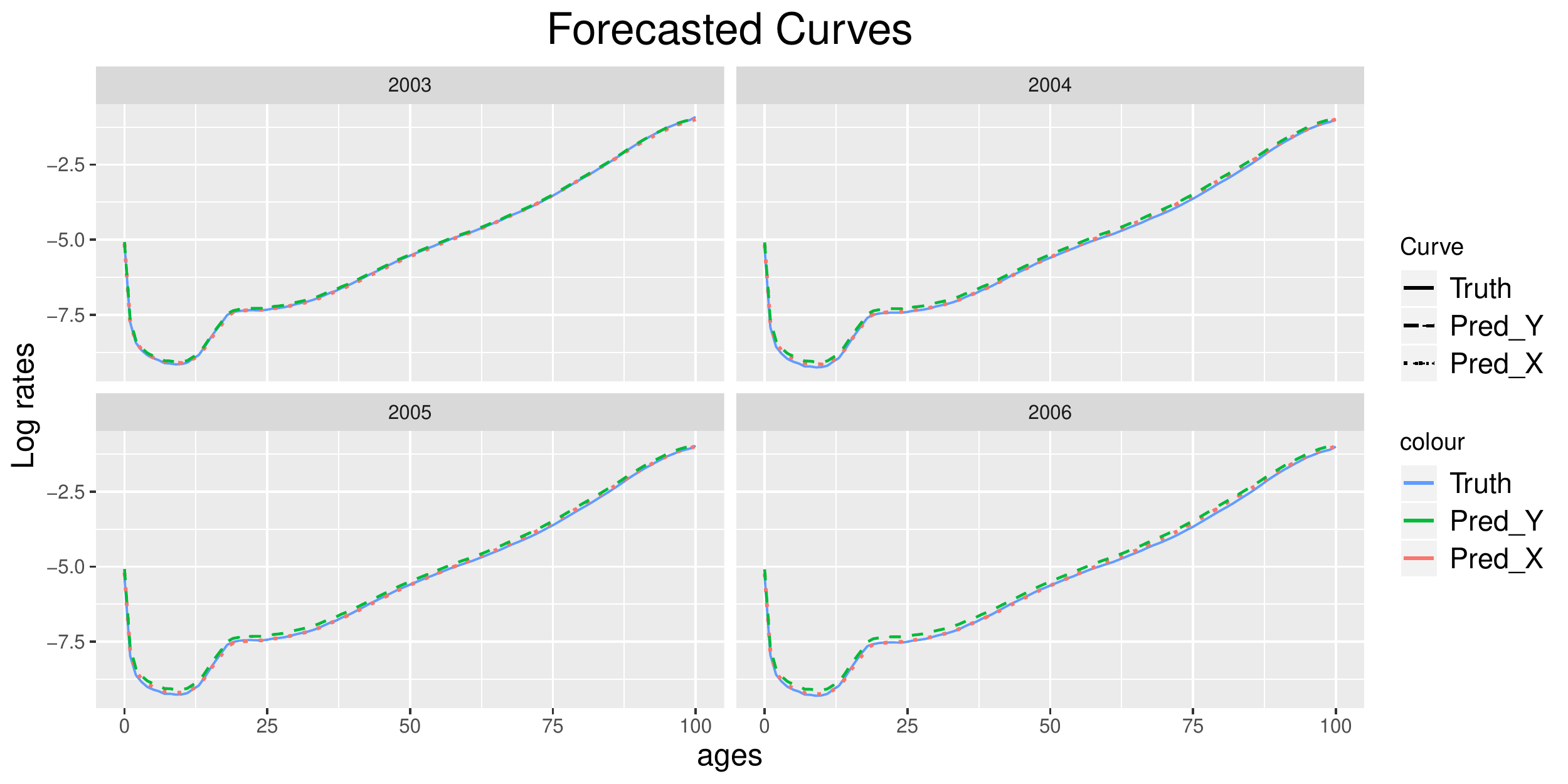}
  \end{center}
    \caption{Forecasting when the functional trend is considered, and when the functional trend is not considered.}\label{Data1Pred}
  \end{subfigure}
  \caption{Results of data analysis. (a) Estimated components of the model \eqref{ModelT}.  (b) Four consecutive curves of log mortality rates with their corresponding forecasted curves. The solid curves (blue) represent the true curves $Y_{2003}(s),\ldots, Y_{2006}$; The dotted curves (red) represent the forecasted curves considering the functional trend, using the time series $\{\tilde{x}_{n,r}\}$; The dashed curves (green) represent the predicted curves without considering the functional trend, using the time series $\{y_{n,r}\}$.}
\end{figure}

This dataset consists of $191$ curves of annual mortality rates in France, from $1816$ to $2006$, for individuals from zero to 100 years old. Each point of the curve $Y_{n}(s)$ represents the log of the mortality rate, in year $n$, at age $s$. At first glance from Figure \ref{Data1} (left), we can say that the functional time series $\{Y_{n}\}$ is nonstationary, and also we can observe a decreasing trend over the years. After applying the stationarity test proposed by \cite{HorvathKokoszkaetal2014}, we obtain a $p$-value equal to $0.003$, and the smaller the $p$-value, the more evidence against the stationarity.  Thus, we consider model \eqref{ModelT}. 

To evaluate the performance of the forecast, we remove  the last $4$ curves of $\{Y_{n}\}$,  that is, we only consider curves from $1816$ to $2002$, with $N=187$.
Figure \ref{Data1} shows the resulting functional time series $Y_{n}$, the estimated functional trend $\hat{T}(s,t)$, and the functional time series $\{\tilde{X}_{n}\}$ after removing the trend (left to right).  We fit ARMA models for the coefficients $\{\tilde{x}_{n,r}, n=1816,\ldots, 2002 \}$, $r=1,2,3,4$. Then, we forecast the $4$ curves $\hat{\tilde{X}}_{2003}, \hat{\tilde{X}}_{2004},  \hat{\tilde{X}}_{2005}$, and $ \hat{\tilde{X}}_{2006}$.  
The models fitted for $\{\tilde{x}_{n,r} \}$ are: ARMA(1,0) with zero mean and coefficient $0.7506$, ARMA(1,0) with zero mean and coefficient $0.9825$, ARMA(1,1) with zero mean and coefficients $(\mathrm{ar}=0.9212, \mathrm{ma}=-0.5593)$, and ARMA(2,0) with zero mean and coefficient $(\mathrm{ar1}= 0.4492, \mathrm{ar2}=0.3601)$, for $r=1,2,3,$ and $4$,  respectively. Also, we forecast the $4$ functional trends $\hat{T}_{2003}, \hat{T}_{2004}, \hat{T}_{2005},$ and $\hat{T}_{2006}$ as described in \eqref{Tforecast}. Finally, we obtain the forecast of the log mortality rate  $\hat{Y}_{2002+h}(s)= \hat{T}_{2002+h}(s) + \hat{\tilde{X}}_{2002+h}(s)$ for $h=1,2,3,4$.

For the case in which the functional trend is not considered, we fit ARIMA models for the coefficients of the projected functional time series, $\{y_{n,r}\}$. In this case the models fitted are: ARIMA$(1,1,1)$ with coefficients $(\mathrm{ar}=0.6562,  \mathrm{ma}=-0.8259, \mathrm{drift}= -0.1213)$, ARIMA$(1,1,1)$ with coefficients $(\mathrm{ar}=0.7606, \mathrm{ma}=-0.9668)$, ARIMA$(1,0,1)$ with coefficients $(\mathrm{ar}=0.8853, \mathrm{ma}=-0.5156)$, ARIMA$(3,1,1)$ with coefficients $(\mathrm{ar1}=0.2569, \mathrm{ar2}= 0.2362,  \mathrm{ar3}=-0.1590, \mathrm{ma}1=-0.6719)$,  for $r=1,2,3,$ and $4$,  respectively. 
We observe that, when the functional trend is not removed, the time series corresponding to the first principal component  $\{y_{n,1}\}$ seems to absorb the trend component. The corresponding time series plots can be found in the  supporting information (Figure $6$).

Figure \ref{Data1Pred} shows the four forecasted curves.  We use different line types and colors to indicate the true curves and   forecasted curves. The solid curves (blue) represent the true curves $Y_{2002+h}(s)$, the dotted curves (red) represent the forecasted curves considering the functional trend, i.e., using the time series $\{\tilde{x}_{n,r}\}$ and forecasting the functional trend, and the dashed curves (green) represent the forecasted curves without considering the functional trend, i.e., using the time series $\{y_{n,r}\}$. Although both methods seem to perform  well, the forecasted curves obtained when considering functional trend are more accurate. Namely, the sum of the $L_{1}$ distance between the truth curves and the predicted curves for each method are $0.449$ and  $0.164$, without/with considering functional trend, respectively.

We observe that the forecasted curves obtained when considering a functional trend are more accurate, i.e., they are closer to the true curves, whereas the forecasted curves obtained when a functional trend is not taken into account are farther away from the true curves. Thus, the consideration of estimating the functional trend improves data analysis.
Based on this, we conclude that the statistical analysis is more accurate when the functional trend is taken into account from the functional point of view.  We recommend estimating such  a functional trend before modeling the stochastic component $\{X_{n}\}$ in model \eqref{ModelT}, either using dimension reduction techniques such as functional principal component, or using a functional time series model such as the functional autoregressive models, FAR$(p)$. 

\section{Discussion}\label{D}
In our study, we assumed a functional time series with a trend component (functional trend). We proposed estimating the functional trend by using a tensor product surface, and taking into account the dependency of the data. To obtain smoothness properties of the estimator, we used marginal penalties. The smoothing parameters were selected based on restricted maximum likelihood, which is robust under correlation structures. We showed that the proposed estimator of the functional trend is consistent when the sample sizes  go to infinity. One of the advantages of our proposal is that it is easy to implement by  using existing R packages, and it can handle large data. In the Monte Carlo simulation, we showed that our functional trend estimator performs well for simple and complex structures of the functional trend.  With the annual mortality rates data, we showed that when the functional trend is  estimated, it improves the inference and the forecasting. 

With this work, we want to encourage taking into account the deterministic component and estimate it from a functional point of view for a functional time series. So, we believe this work will be of interest for data applications. Also, this work leads to a future project that is the extension to functional time series with domain in $\mathbb{R}^{2}$, called surface time series \citep[][]{MARTINEZHERNANDEZ2020}. Such an extension could benefit, for example, fMRI data and spatio-temporal data in general.

\end{document}